\documentclass[aps,preprint,nofootinbib]{revtex4}
\usepackage{amssymb}
\usepackage{graphicx}
\usepackage{epsfig}

\begin{document}

%=====================================
%       put your definitions here
%=====================================
\def\beq{\begin{eqnarray}}
\def\eeq{\end{eqnarray}}
\def\non{\nonumber}
\def\la{\langle}
\def\ra{\rangle}
\def\Un{{\cal U}}
\def\Mbar{\overline{M^0}}
\def\Bmixing{B^0-\overline{B^0}}
\def\Dmixing{D^0-\overline{D^0}}

\def\pr{{Phys. Rev.}~}
\def\prl{{ Phys. Rev. Lett.}~}
\def\pl{{ Phys. Lett.}~}
\def\npb{{ Nucl. Phys. B}~}
\def\epjc{{ Eur. Phys. J. C}~}

%======================================================================

\title{ Neutrino decay as a possible interpretation to the MiniBooNE observation
 with unparticle scenario }

\author{Xue-Qian Li$^{1}$}
\author{Yong Liu$^{2}$}
\author{Zheng-Tao Wei$^{1}$}
\author{Liang Tang$^{1}$}

\affiliation{$^{1}$ Department of Physics, Nankai University,
  Tianjin 300071, China}
\affiliation{$^{2}$ University of Alabama, Tuscaloosa, AL 35487}

\begin{abstract}

\noindent In a new measurement on neutrino oscillation
$\nu_{\mu}\to\nu_e$, the MiniBooNE Collaboration observes an excess
of electron-like events at low energy and the phenomenon may demand
an explanation which obviously is beyond the oscillation picuture.
We propose that heavier neutrino $\nu_2$ decaying into a lighter one
$\nu_1$ via the transition process $\nu_{\mu}\to \nu_e+X$ where $X$
denotes any light products, could be a natural mechanism. The
theoretical model we employ here is the unparticle scenario
established by Georgi. We have studied two particular modes
$\nu_\mu\to \nu_e+\Un$ and $\nu_\mu\to \nu_e+\bar\nu_e+\nu_e$.
Unfortunately, the number coming out from the computation is too
small to explain the observation. Moreover, our results are
consistent with the cosmology constraint on the neutrino lifetime
and the theoretical estimation made by other groups, therefore we
can conclude that even though neutrino decay seems plausible in this
case, it indeed cannot be the source of the peak at lower energy
observed by the MiniBooNE collaboration and there should be other
mechanisms responsible for the phenomenon.

\end{abstract}

\maketitle

\section{Introduction}

Recently, the MiniBooNE Collaboration reported its results of
searching for $\nu_\mu\to\nu_e$ oscillations \cite{MiniBooNE}. In
the experiment, the $\nu_\mu$ energy spectrum has a peak centered
at 700 MeV and extends to 3000 MeV. For the oscillation range
$475<E_{\nu}<1250$ MeV where $E_{\nu}$ is the energy of the
produced neutrino, no significant excess of events is found. This
result excludes sizable appearance of $\nu_e$ via two neutrino
oscillation and disfavors the previous LSND measurement
\cite{LSND}. However, they observed that outside of the
oscillation range, there is a clear peak of the
electron-neutrino-like events ($96\pm 17\pm 20$ events) lying
above background  at $300<E_{\nu}<475$ MeV. Although the origin of
the excess is still under investigation, we may assume that they
are indeed electron neutrinos at present. The beam is completely
composed of $\nu_{\mu}$ and the oscillation can only produce
$\nu_e$ with the same energy, therefore the observation would be a
serious challenge to the present theories. Namely a reasonable
explanation about the appearance of the low energy $\nu_e$ is
needed. To answer this question,  there are some interesting
proposals, for example, in \cite{MS} the authors suggest a (3+2)
neutrino oscillation scenario where two sterile neutrinos are
introduced into the game to explain the MiniBooNE results and
Bodek\cite{Bodek} considered the internal bremsstrahlung as an
alternative source of the excess $\nu_e$ events. Instead, in this
work, we are looking for possible mechanisms other than the
neutrino oscillation, supposing that there are only standard model
(SM) neutrinos.  An explanation that $\nu_{\mu}$ may decay into
$\nu_e+X$ where $X$ denotes some possible light products, seems
reasonable. Definitely, neutrino decay must be realized via
interactions beyond the SM. The possible candidates of $X$ could
be $\nu_e+\bar\nu_e$, light bosons (for example axion etc. ) and
the unparticle which we are going to explore in this work.

In fact, the idea of neutrino decay is not new. It has been put
forward by some authors \cite{MRS,PPS}. The basic idea is to
introduce a heavy, unstable neutrino (usually assuming a sterile
one) which decays into light neutrino or antineutrino plus a
scalar particle. The interactions between the scalar particle and
neutrinos are described by a lepton flavor violating effective
lagrangian which depends on the details of various new physics
models. Instead, we suggest an alternative scenario, namely the
heavier $\nu_2$ which is a mass eigenstate and a component of the
flavor eigenstate $\nu_{\mu}$, is the constituents of the beam and
decays into a light neutrino $\nu_1$ and a scale-invariant
unparticle proposed recently by Georgi \cite{Georgi1}.

It is well known that at very high energy scale, the unparticle
physics contains the SM fields and a sector of Banks-Zaks field
(defined in \cite{BZ}) with a non-trivial infrared fix point. Below
an energy scale $\Lambda_\Un$ which is of order of TeV, the
Banks-Zaks fields are matched onto a scale invariant unparticle
sector. The unparticle is different from the ordinary particles as
it has no mass since the mass term breaks the scale invariance, but
the Lorentz-invariant four-momentum square needs not to be zero,
$P^2\geq 0$. The scale dimension of unparticle is in general
fractional rather than an integral number (the dimension for a
fermion is half-integers). This special characteristic brings us a
natural explanation of the shape of low energy $\nu_e$ bump observed
by the MiniBooNE Collaboration. If $\nu_{2}$ decays into a $\nu_1$
and a real scalar particle where $\nu_i$ are neutrino mass
eigenstates, it is a two-body decay where the energy-spectrum of the
produced $\nu_e$ should be discrete. It is in contrary to the
observation where the energy spectrum of the produced $\nu_e$ is
continuous. Indeed, the incident $\nu_{\mu}$ beam has an energy
distribution which can result in a natural energy spreading for the
produced electron-neutrino, however, it demands that the shape of
$\nu_e$ spectrum must be similar to that of the incident $\nu_{\mu}$
beam. Instead, if the produced $X$ is an unparticle, the energy
spectrum of $\nu_e$ would naturally spread and it may be more
consistent with the present measurements. The interactions between
the unparticle and the SM particles are described in the framework
of low energy effective theory and lead to various interesting
phenomenology. There have been many phenomenological explorations on
possible observable effects caused by unparticles
\cite{Georgi1,Georgi2,CKY,LZ,CG1,Liao,DY1,
ACG1,LW,LWW,Ste,FRS,Gre,Dav,CGM,CH,ACG2,MR,Zhou,DY2,CG2,LL2,BFRS,Rizzo,
CKY2,GN,CHT,Zwicky,KO,MG,HW,Lenz,CGF} and much more are coming up.

The MiniBooNE results indicate that the energy of the events of
excess is about a half of the peak position at the energy spectrum
of the muon neutrino. As discussed above, we suggest a decay mode
$\nu_2\to\nu_1+\Un$ where $\Un$ denotes the unparticle and a
consequent transition $\nu_{\mu}\to\nu_e+\Un$ might be observed,
namely $\nu_{\mu}$ and $\nu_e$ are not physical eigenstates, but are
that of weak interaction and can be caught by detector as an
appearance of $\nu_e$ at lower energy. As indicated in
\cite{Georgi1}, the unparticle stuff with scale dimension $d_\Un$
cannot be "seen" directly, it would manifest itself as a missing
energy. When the scale dimension $d_\Un$ is not very large, the
energy spectrum of electron neutrino can fall into the allowed range
of the MiniBooNE measurements. This process has also been considered
in \cite{Zhou}. A transition into a three-body final state
$\nu_\mu\to\nu_e+\nu_e+\bar \nu_e$ is also a possible process to
explain the MiniBooNE data. The two decay modes:
$\nu_\mu\to\nu_e+\Un$ and $\nu_\mu\to\nu_e+\nu_e+\bar \nu_e$ are
both lepton flavor violating processes and can only occur via new
physics mechanism beyond the SM.

If the mechanism proposed above can explain the observed peak which
depends on its decay width, the possible detection rate of the
number of $\nu_e$ is roughly
\begin{equation}
N_{\nu_e}\sim N_0\left(1-e^{-t/\tau}\right)\times \eta,
\end{equation}
where $N_0$ is the muon neutrino number, $\tau$ is the lifetime
which should be calculated in the aforementioned scenario and $\eta$
is a detection rate which is also very small, say, $10^{-10}$ or
even smaller. t is the flight time from the source to the detector
and since the speed of the beam neutrino is very close to the speed
of light, $t\sim L/c$ where $L$ is the distance and approximately
500 m in the MiniBooNE experiments. In addition the ratio would be
further suppressed by the time dilation factor $\gamma=m/E$. Since
$L$ is only of several hundred meters, to make a sizable ratio which
can be observed, $\tau$ must not be large. We will obtain its value
by imputing all the concerned model parameters which are fixed by
fitting data of other experiments into our numerical computation.

As is well-known, neutrino oscillation had been observed in the
solar, atmospheric, accelerator neutrino experiments and the present
theoretical studies almost completely confirm the MSW mechanism. The
relevant mixing parameters and the mass-square differences are
determined by fitting the data, even though the absolute values of
the neutrino masses are still not fixed yet. While theoretically
determining the parameters, possible neutrino decays are not taken
into account seriously. How to reconcile the neutrino decay with the
present theoretical works on the neutrino oscillation is an open
question, namely, one should explore if there exists discrepancy
between the theoretical predictions and data. Obviously if the decay
rate is sufficiently small, one does not need to modify the present
theoretical framework about the neutrino oscillation, but if the
decay rate is not too small, the data fitting should be
re-considered, therefore the MiniBooNE result indeed provides a new
challenge to the neutrino physics and we will return to this topic
in our next work \cite{Li}.

\section{ Neutrino decays in unparticle physics}

We start with a brief review of the unparticle physics. First let
us consider an unparticle $\Un$ with scale dimension $d_\Un$ and
momentum $P$. The unparticle momentum satisfies the constraint
$P^2\geq 0$. According to \cite{Ste}, the unparticle stuff can be
viewed as a tower of massive particles with mass spacing tends to
zero. Scale invariance provides the most important constraint on
the properties of unparticles. The two-point function of scalar
unparticle field operator $O_\Un$ is written as
 \beq
 \la 0|O_\Un(x) O^\dagger_\Un(0)|0\ra=\int \frac{d^4 P}{(2\pi)^4}e^{-iP\cdot x}
  |\la0|O_\Un(0)|P\ra|^2\rho(P^2),
 \eeq
where $|P\ra$ is the unparticle state with momentum $P$ and the
phase space factor is
 \beq
 |\la0|O_\Un(0)|P\ra|^2\rho(P^2)=
  A_{d_\Un}\theta(P^0)\theta(P^2)(P^2)^{d_\Un-2},
 \eeq
where
 \beq
 A_{d_\Un}=\frac{16\pi^{5/2}}{(2\pi)^{2d_\Un}}\frac{\Gamma(d_\Un+1/2)}
  {\Gamma(d_\Un-1)\Gamma(2d_\Un)}.
 \eeq
For the vector unparticle field $O_\Un^{\mu}$, we have
 \beq
 \la0|O_\Un^{\mu}(0)|P\ra\la P|O_\Un^{\nu}(0)|0\ra\rho(P^2)=
  A_{d_\Un}\theta(P^0)\theta(P^2)(-g^{\mu\nu}+P^\mu
  P^\nu/P^2)(P^2)^{d_\Un-2},
 \eeq
where the transverse condition $\partial_\mu O_\Un^\mu=0$ is
required. The Lorentz structure of unparticle can also be tensor
\cite{CKY,CKY2} or spinor \cite{LZ}. In this study, we restrict our
discussions to scalar and vector unparticles. Obviously similar
analysis can be done for the tensor and spinor unparticles.

About the virtual effects, the propagator of the scalar unparticle
field  is given as
 \beq
 \int d^4 x e^{iP\cdot x}\la 0 |TO_\Un(x)O_\Un(0)|0\ra &=&
   i\frac{A_{d_\Un}}{2{\rm sin}(d_\Un\pi)}\frac{1}{(P^2+i\epsilon)^{2-d_\Un}}
   e^{-i(d_\Un-2)\pi},
 \eeq
and for the vector unparticle field, the propagator is
 \beq
 \int d^4 x e^{iP\cdot x}\la 0 |TO^{\mu}_\Un(x)O^{\nu}_\Un(0)|0\ra &=&
  i\frac{A_{d_\Un}}{2{\rm sin}(d_\Un\pi)}\frac{-g^{\mu\nu}+P^{\mu}P^{\nu}/P^2}
  {(P^2+i\epsilon)^{2-d_\Un}}e^{-i(d_\Un-2)\pi}.
 \eeq
The function ${\rm sin}(d_\Un\pi)$ at the denominator implies that
the scale dimension $d_\Un$ cannot be integers for $d_\Un>1$ in
order to avoid singularity. The phase factor $e^{-i(d_\Un-2)\pi}$
provides a CP conserving phase which produces peculiar
interference effects in high energy scattering processes
\cite{Georgi2,CKY,CKY2} and CP violation in B decays
\cite{CG1,CG2}.

In this study, we will discuss interactions between the unparticles
and neutrinos. The framework which describes these interactions is a
low energy effective theory. For our purpose, the coupling of
unparticle to neutrinos ($\nu_\mu$ and $\nu_e$) is given in the form
as
 \beq
 {\cal L}_{eff}=\frac{c_S^{\nu_\alpha\nu_\beta}}{\Lambda_\Un^{d_\Un}}\bar
  \nu_\beta\gamma_{\mu}(1-\gamma_5)\nu_\alpha\partial^\mu O_\Un
 +\frac{c_V^{\nu_\alpha\nu_\beta}}{\Lambda_\Un^{d_\Un-1}}\bar
  \nu_\beta\gamma_{\mu}(1-\gamma_5)\nu_\alpha O_\Un^\mu +h.c. .
 \eeq
Here, we have used the $V-A$ type current as in the SM. The $c_S$
and $c_V$ are dimensionless coefficients. The $\nu_\alpha$ and
$\nu_\beta$ are weak eigenstates with different flavor numbers
$\alpha$ and $\beta$.

As in \cite{Zhou}, the neutrino decay is conveniently represented
in the basis of mass eigenstates $\nu_i$ (i=1,2, we only consider
two generations in this case). The interactions between unparticle
and neutrinos are rewritten by
 \beq
 \frac{c_S^{\nu_i\nu_j}}{\Lambda_\Un^{d_\Un}}\bar
  \nu_j\gamma_{\mu}(1-\gamma_5)\nu_i\partial^\mu O_\Un
 +\frac{c_V^{\nu_i\nu_j}}{\Lambda_\Un^{d_\Un-1}}\bar
  \nu_j\gamma_{\mu}(1-\gamma_5)\nu_i O_\Un^\mu +h.c. .
 \eeq
The relation between the coupling coefficients
$c_{S(V)}^{\nu_\alpha\nu_\beta}$ and $c_{S(V)}^{\nu_i\nu_j}$ can be
obtained from neutrino mixing matrix. For a simple case, considering
two neutrino mixing,
 \beq
 \left(  \begin{array}{c} \nu_e   \\
  \nu_{\mu}  \end{array} \right)=
 \left( \begin{array}{cc} {\rm cos}\theta & -{\rm sin}\theta  \\
   {\rm sin}\theta &  {\rm cos}\theta \end{array} \right)
 \left( \begin{array}{c} \nu_1 \\
  \nu_2  \end{array} \right),
 \eeq
The coefficients in the mass basis are related to those in the
flavor basis by
 \beq
 c_{S(V)}^{\nu_1\nu_2}={\rm cos}^2\theta~ c_{S(V)}^{\nu_\alpha\nu_\beta}.
 \eeq
For a maximal mixing where $\theta=\pi/4$,
$c_{S(V)}^{\nu_1\nu_2}=\frac{1}{2}~ c_{S(V)}^{\nu_\alpha\nu_\beta}$.
The coefficients in the different basis differ by a constant factor.

\subsection{The decay of $\nu_2\to\nu_1+\Un$}

\begin{figure}[!htb]
\begin{center}
\begin{tabular}{cc}
\includegraphics[width=4cm]{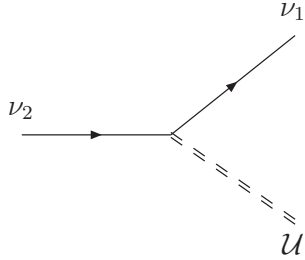}
\end{tabular}
\end{center}
\caption{ The diagram for the decay of $\nu_2\to\nu_1+\Un$. The
double dashed lines represent the unparticle.} \label{fig1}
\end{figure}

Assuming two generation neutrinos and the heavier one is $\nu_2$ and
lighter is $\nu_1$, the decay of $\nu_\mu\to\nu_e+\Un$ is realized
via the transition $\nu_2\to\nu_1+\Un$ which is a typical lepton
flavor violating process and its Feynman diagram is depicted in Fig.
\ref{fig1}. Here, it is natural to assume that the basis of the
interaction between unparticle and neutrino is the same as the weak
interaction, namely $\nu_e$ and $\nu_{\mu}$ are the eigenstates of
the interaction. The final unparticle is invisible and behaves as a
missing energy. The decay $\nu_2\to\nu_1+\Un$ seems to be a two-body
process. But it is different from the common case with two final
particles whose momenta are single-valued and fixed. For the
unparticle case, the energy of $\nu_1$ depends on the momentum
square of unparticle $P^2$ which only is constrained by the
condition $P^2\geq 0$. Namely, one can expect that $P^2$ can vary
within a range $0\leq P^2\leq P_{max}^2$ where $P_{max}$ would be
determined by the momenutum conservation in $\nu_{\mu}$ decay. Thus
the varying $P^2$ causes a continuous energy spectrum of $E_{\nu_e}$
and it is a characteristic effect of the unparticle.

In order to make the process realizable, the mass of $\nu_2$ should
be larger than that of $\nu_1$, that is the so-called normal order
in literature. Without losing generality, we further assume
$m_{\nu_2}\gg m_{\nu_1}$ and neglect $m_{\nu_1}$ in the analysis
below. The differential decay rate of
$\nu_2(p_2)\to\nu_1(p_1)+\Un(q)$ is
 \beq
 d\Gamma=\frac{1}{2E_{\nu_2}}\frac{1}{2}\sum_{\rm spins}|{\cal M}|^2~d\Phi(p),
 \eeq
where the phase space factor $d\Phi(p)$ is
 \beq
 d\Phi(p)=(2\pi)^4\delta^4(p_2-p_1-q)\left[ 2\pi\theta(p_1^0)\delta(p_1^2)
  \frac{d^4p_1}{(2\pi)^4}\right]\left[A_{d_\Un}\theta(q^0)\theta(q^2)
  (q^2)^{d_\Un-2}\frac{d^4q}{(2\pi)^4}\right],
 \eeq
with $q=p_2-p_1$ and the Lorentz-invariant amplitude square is
 \beq
 \frac{1}{2}\sum_{\rm spins}|{\cal M}|^2=\frac{|c_S^{\nu_1\nu_2}|^2}
  {\Lambda_\Un^{2d_\Un}}~4~p_2^2(p_1\cdot p_2),
 \eeq
for the scalar unparticle and
 \beq
 \frac{1}{2}\sum_{\rm spins}|{\cal M}|^2=\frac{|c_V^{\nu_1\nu_2}|^2}
  {\Lambda_\Un^{2d_\Un-2}}~4\left[2(p_1\cdot p_2)+
  \frac{p_2^2(p_1\cdot p_2)}{q^2}\right].
 \eeq
for the vector unparticle.

In the rest frame of $\nu_2$, it is straightforward to derive the
differential decay rate over the $\nu_1$ energy $E_1$ and the decay
rate of $\nu_2(p_2)\to\nu_1(p_1)+\Un(q)$ as
 \beq \label{restS}
 \frac{d\Gamma_S}{dE_1}&=&\frac{|c_S^{\nu_1\nu_2}|^2A_{d_\Un}}
  {2\pi^2\Lambda_\Un^{2d_\Un}}\frac{m_{\nu_2}^2E_1^2}
  {(m_{\nu_2}^2-2m_{\nu_2}E_1)^{2-d_\Un}}\theta(m_{\nu_2}-2E_1), \non\\
 \Gamma_S&=&\frac{|c_S^{\nu_2\nu_1}|^2A_{d_\Un}}{8\pi^2d_\Un(d_\Un^2-1)}
   m_{\nu_2}\left(\frac{m_{\nu_2}}{\Lambda_\Un}\right)^{2d_\Un},\\
 \frac{d\Gamma_V}{dE_1}&=&\frac{|c_V^{\nu_1\nu_2}|^2A_{d_\Un}}
  {2\pi^2\Lambda_\Un^{2d_\Un-2}}\frac{m_{\nu_2}^2E_1^2
  \left[ 1+2\left(1-2\frac{E_1}{m_{\nu_2}}\right) \right]}
  {(m_{\nu_2}^2-2m_{\nu_2}E_1)^{3-d_\Un}}\theta(m_{\nu_2}-2E_1),\non\\
 \Gamma_V&=&\frac{3|c_V^{\nu_1\nu_2}|^2A_{d_\Un}}{8\pi^2d_\Un(d_\Un+1)(d_\Un-2)}
   m_{\nu_2}\left(\frac{m_{\nu_2}}{\Lambda_\Un}\right)^{2d_\Un-2}.
   \label{restV}
 \eeq
with $d_\Un>1$ for the scalar unparticle and $d_\Un>2$ for the
vector unparticle.

In this study, we concern the observation in the laboratory frame
where the initial muon neutrino moves nearly with the speed of
light. The energy of $\nu_2$ is at the order of several hundred
MeV and is much bigger than its inertial mass, $E_2\gg m_{\nu_2}$.
The invariant amplitude square in the laboratory frame depends on
the angle $\theta'$ between the moving directions of muon and
electron neutrinos. Indeed, we need to do some treatments to get
an analytical result, i.e. boost the result in the rest frame of
$\nu_2$ into the laboratory frame. The momentum of $\nu_2$ is
approximated by $|\vec p_2|=\sqrt{E_2^2-m_{\nu_2}^2}\cong
E_2(1-\frac{m_{\nu_2}^2}{2E_2^2})$ and the momentum product
$p_2\cdot p_1\cong E_1E_2(\frac{m_{\nu_2}^2}{2E_2^2}+1-{\rm
cos}\theta')$. From $q^2\geq 0$, the range of ${\rm cos}\theta'$
is determined to be $0\leq 1-{\rm cos}\theta'\leq
\frac{m_{\nu_2}^2}{2E_2 E_1}(1-\frac{E_2}{E_1})$ which means that
three-momentum of the produced electron neutrino is almost
parallel to that of the muon neutrino. After performing an
integration over ${\rm cos}\theta'$, we obtain the differential
decay rate of $\nu_2(p_2)\to\nu_1(p_1)+\Un(q)$ in the laboratory
frame as
 \beq \label{e1S}
 \frac{d\Gamma_S}{dE_1}&=&\frac{|c_S^{\nu_1\nu_2}|^2A_{d_\Un}}
  {4\pi^2}\left(\frac{m_{\nu_2}^2}{\Lambda_\Un^2}\right)^{d_\Un}
  \frac{m_{\nu_2}^2}{E_\mu^2}\frac{\left[1+y(d_\Un-1)\right]
  \left(1-y\right)^{d_\Un-1}}{d_\Un(d_\Un-1)}\theta(E_2-E_1),
 \eeq
for the scalar unparticle with $y\equiv\frac{E_1}{E_2}$ and
$d_\Un>1$. For the vector unparticle, the differential decay rate is
 \beq \label{e1V}
 &&\frac{d\Gamma_V}{dE_1}=\frac{|c_V^{\nu_1\nu_2}|^2A_{d_\Un}}
  {16\pi^2}\left(\frac{m_{\nu_2}^2}{\Lambda_\Un^2}\right)^{d_\Un-1}
  \frac{m_{\nu_2}^2}{E_2^2}(1-y)^{d_\Un-2}\Gamma(d_\Un+1) \non\\
 &&~~~~~~~~\times\left[
          y(3-2y) ~\frac{_2F_1(1,3;d_\Un+2;1)}{\Gamma(d_\Un+2)}
      +(3-7y+4y^2)~\frac{_2F_1(2,3;d_\Un+3;1)}{\Gamma(d_\Un+3)}\right.
     \non\\
 &&~~~~~~~~~~~~~\left.
         -4(1-y)^2~\frac{_2F_1(3,3;d_\Un+4;1)}{\Gamma(d_\Un+4)}\right]
  \theta(E_2-E_1).
 \eeq
with $d_\Un>2$ and $_2F_1(a,b;c;z)$ is the hypergeometric function.
The decay rates $\Gamma_S$ and $\Gamma_V$ can be obtained and the
final results differ from Eqs. (\ref{restS}) and (\ref{restV}) by a
Lorentz factor $\frac{m_{\nu_2}}{E_2}$.

\subsection{The three-body decay of $\nu_2\to\nu_1+\bar\nu_1+\nu_1$}

\begin{figure}[!htb]
\begin{center}
\begin{tabular}{cc}
\includegraphics[width=14cm]{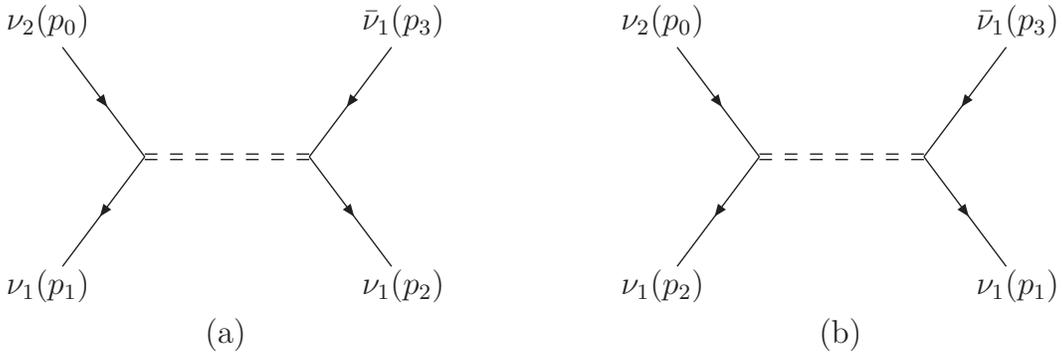}
\end{tabular}
\end{center}
\caption{ The Feynman diagram for the three-body decay of $\nu_2\to
\nu_1+\nu_1+\bar\nu_1$. } \label{fig2}
\end{figure}

As briefly discussed in the introduction, there is another
possibility to observe a continuous energy spectrum of $\nu_{1}$.
Now, we turn to the three-body decay of $\nu_{2}$ in the framework
of unparticle: $\nu_2\to\nu_1+\bar\nu_1+\nu_1$. The Feynman diagram
is depicted in Fig. \ref{fig2} where the unpartilce serves as an
intermediate agent. Because the final states have two electron
neutrinos, there are two diagrams in the process and one needs to
consider the anti-symmetrization of the two identical fermions. We
consider only the vector unparticle part because the scalar
unparticle contribution is proportional to the light neutrino mass
and should be very suppressed. According to the effective
interaction of neutrinos and unparticle, the decay amplitude of
$\nu_2(p_0)\to\nu_1(p_1)+\nu_1(p_2)+\bar\nu_1(p_3)$ is
 \beq
 {\cal M}&=&{\cal M}_1+{\cal M}_2, \non\\
 {\cal M}_1&=&-\frac{c_V^{\nu_1\nu_2}c_V^{\nu_1\nu_1}}{\Lambda_\Un^{2d_\Un-2}}
  ~\frac{A_{d_\Un}e^{-i\phi}}{2{\rm sin}d_\Un\pi}~\frac{\bar u(p_1)\gamma_\mu
  (1-\gamma_5)u(p_0)\bar u(p_2)\gamma_\mu(1-\gamma_5)v(p_3)}
  {(q_1^2)^{2-2d_\Un}}, \non \\
 {\cal M}_1&=&+\frac{c_V^{\nu_1\nu_2}c_V^{\nu_1\nu_1}}{\Lambda_\Un^{2d_\Un-2}}
  ~\frac{A_{d_\Un}e^{-i\phi}}{2{\rm sin}d_\Un\pi}~\frac{\bar u(p_2)\gamma_\mu
  (1-\gamma_5)u(p_0)\bar u(p_1)\gamma_\mu(1-\gamma_5)v(p_3)}
  {(q_2^2)^{2-2d_\Un}},
 \eeq
where $\phi=(d_\Un-2)\pi$, $q_1=p_0-p_1$ and $q_2=p_0-p_2$. The
amplitudes ${\cal M}_1$ and ${\cal M}_2$ represent contributions
from Fig. \ref{fig2}(a) and (b), respectively. In the derivations,
we have neglected $q_1^\mu q_1^\nu/q_1^2$ and $q_2^\mu
q_2^\nu/q_2^2$ terms. The square of the invariant matrix element is
 \beq
 \frac{1}{2}\sum_{\rm spins}|{\cal M}|^2=\frac{|c_V^{\nu_1\nu_2}c_V^{\nu_1\nu_1}|^2}
  {8\Lambda_\Un^{4d_\Un-4}}~\frac{A_{d_\Un}^2}{4({\rm sin}d_\Un\pi)^2}~
  \left[\frac{1}{(q_1^2)^{2-2d_\Un}}+\frac{1}{(q_2^2)^{2-2d_\Un}}\right]^2
  256(p_0\cdot p_3)(p_1\cdot p_2).
 \eeq

For the three-body decays, we work in the rest frame of the muon
neutrino. As will be shown later, the three-body decay rate of
$\nu_2\to\nu_1+\bar\nu_1+\nu_1$ is much smaller than that of the
two-body decay $\nu_2\to\nu_1+\Un$. This conclusion does not
depend on which reference frame we choose. In the rest frame of
$\nu_2$, the three-body kinematics are described in terms of the
final neutrino energies $E_1$, $E_2$ by
 \beq
 &&p_0\cdot p_3=m_{\nu_\mu}\left(m_{\nu_\mu}-E_1-E_2\right), \qquad
 ~~p_1\cdot p_2=m_{\nu_\mu}\left(E_1+E_2-\frac{m_{\nu_\mu}}{2}\right),\non\\
 &&q_1^2=(p_0-p_1)^2=m_{\nu_\mu}^2-2m_{\nu_\mu}E_1, \qquad
 q_2^2=(p_0-p_2)^2=m_{\nu_\mu}^2-2m_{\nu_\mu}E_2,
 \eeq
Thus, the differential decay rate is
 \beq
 d\Gamma_V(\nu_2\to \nu_1+\nu_1+\bar\nu_1)=\frac{1}{(2\pi)^3}
  \frac{1}{8m_{\nu_2}}~\frac{1}{2}~\frac{1}{2}|{\cal M}|^2
  ~dE_1dE_2.
 \eeq
where the integration range is $0\leq E_1\leq \frac{m_{\nu_2}}{2}$
and $\frac{m_{\nu_2}}{2}-E_1\leq E_2\leq \frac{m_{\nu_2}}{2}$.

Note that the similar lepton flavor violating processes $\mu^-\to
e^-+\Un$, $\mu^-\to e^-+e^-+e^+$ have been studied in
\cite{ACG1,CGM} and their formulations are quite similar to ours.

\section{Analysis on the concerned phenomenology}

For the neutrino accelerator experiment, the neutrinos fly over a
baseline with distance $L$ before reaching at the final detectors.
If neutrino decays as we suggested, the number of final electron
neutrinos produced from muon neutrino decaying is
 \beq
 N_{\nu_e}=N_0~{\rm exp}\left[1-\left(-\frac{t}{\tau_{\rm lab}}\right)\right]
  \approx N_0~\frac{L}{c~\tau_\nu}~\frac{m}{E}.
 \eeq
with $N_0$ the initial muon neutrino number, $\tau_{\rm lab}$ and
$\tau_\nu$ are neutrino life times in the laboratory and rest frame,
respectively. In the MiniBooNE measurement, $L/E\sim 500~{\rm
m}/500~{\rm MeV}$, $N_{\nu_e}\sim 100$. The ratio of neutrino life
time over mass $\tau_\nu/m_\nu\sim \frac{N_0}{N_{\nu_e}}
10^{-14}~{\rm s/eV}$. When $N_0/N_{\nu_e}\sim 10^{10}$,
$\tau_\nu/m_\nu\sim 10^{-4}$, and when $N_0/N_{\nu_e}\sim 10^{5}$,
$\tau_\nu/m_\nu\sim 10^{-9}$.

At present, our knowledge on the neutrino mass is mainly obtained
from the neutrino oscillation data. The squared mass difference of
the mass eigenstates $\Delta m^2_{ij}\equiv m_i^2-m_j^2$ is observed
to be \cite{PDG}: $\Delta m^2\sim 8\times 10^{-5} {\rm eV}^2$,
$\Delta m^2\sim 3\times 10^{-3} {\rm eV}^2$. We will not use the
LSND result ($\Delta m^2\sim 1{\rm eV}^2$) since it is disfavored by
other neutrino experiments and the new MiniBooNE measurements. From
these data, we choose $m_2=50$ meV as the upper limit in our
numerical calculations.

Firstly, we give an observation that the rate of three-body decay
$\nu_\mu\to\nu_e+\bar\nu_e+\nu_e$ is much smaller than two-body
process $\nu_\mu\to \nu_e+\Un$. For an illustration, we choose the
parameters as $c_S=c_V=1$, $\Lambda_\Un=1$ TeV, $d_\Un=1.1$ for
scalar and $d_\Un=2.1$ for vector unparticles. The parameters
satisfy the cosmological constraints. For the vector unparticle
contribution, the rates of two-body and three-body decays are
 \beq
 &&\Gamma_V(\nu_\mu\to\nu_e+\Un)=4\times 10^{-34}~{\rm eV}, \non\\
 &&\Gamma_V(\nu_\mu\to\nu_e+\bar\nu_e+\nu_e)=1\times 10^{-66}~{\rm eV}.
 \eeq
The three-body decay rate is more than 30 orders smaller than that
of the two-body case. The tiny ratio is due to the very weak
coupling between unparticle and neutrinos (and there are two such
vertices for the process, see Fig. 2.) and small neutrino mass. This
observation is analogous to the process of $\mu^-\to
e^-+\nu_\mu+\bar\nu_e$ where the decay rate is proportional to
$G_F^2$ and $m_\mu^5$. If we only use the three-body decay, a life
time of neutrino which is so long that muon neutrino will never
decay and the bump observed in the experiment cannot be explained by
the neutrino decays at all. Thus, we can safely neglect the
contributions from the three-body decays and approximate
$\Gamma_{\nu}=\Gamma(\nu_\mu\to\nu_e+\Un)$.

Secondly, we discuss the constraints of unparticle parameters from
neutrino decays. As discussed above, the MiniBooNE experiments put a
bound for neutrino life time in the rest frame $\tau_\nu/m_\nu\sim
10^{-4}~{\rm s/eV}$. We take this bound for our analysis and discuss
three possibilities. In order for the illustration, we are
restricted in the case of vector unparticle. (1) We fix $d_\Un=2.1$,
$\Lambda_\Un=1$ TeV, and constrain $c_V$ by $c_V>10^{11}$. The
coupling constants are found to be much larger than the order 1. (2)
We fix $c_S=c_V=1$,  $\Lambda_\Un=1$ TeV, and constrain
$0<d_\Un-2<10^{-7}$. The scale dimension will be nearly equal to 2.
(3) We fix $c_S=c_V=1$, $d_\Un=2.1$, and constrain
$\Lambda_\Un<10^{-9}$ TeV which is obviously impossible. Thus, if
neutrino decays as suggested, the unparticle parameters have to fall
into a very unnatural space. On the opposite side, if the unparticle
parameters are chosen in a reasonable space, the neutrino life time
will be so long that they will not decay when flying over the
distance $L\sim 500$m in the MiniBooNE.

\begin{figure}[!htb]
\begin{center}
\begin{tabular}{cc}
\includegraphics[width=8cm]{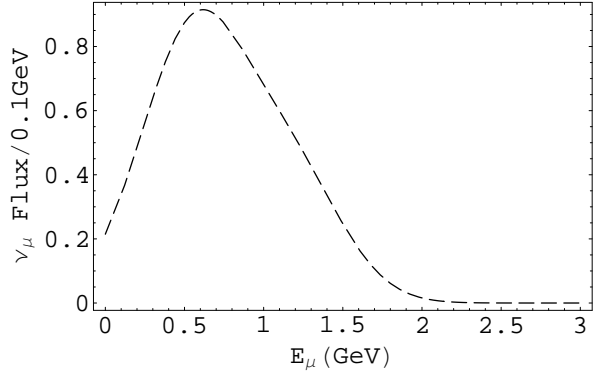}
\end{tabular}
\end{center}
\caption{ The energy spectrum for the $\nu_\mu$ beam in the
laboratory frame. } \label{fig4}
\end{figure}

\begin{figure}[!htb]
\begin{center}
\begin{tabular}{cc}
\includegraphics[width=10cm]{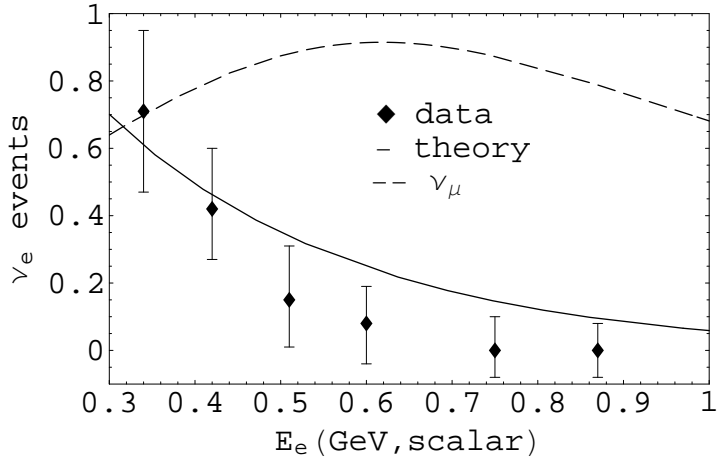}
\end{tabular}
\end{center}
\caption{ The energy spectrum for the decay of $\nu_\mu\to
\nu_e+\Un$ with scalar unparticle in the laboratory frame where
$d_\Un=1.1$, $\Lambda_\Un=1$ TeV and $c_S=c_V=1$. } \label{fig5}
\end{figure}

\begin{figure}[!htb]
\begin{center}
\begin{tabular}{cc}
\includegraphics[width=10cm]{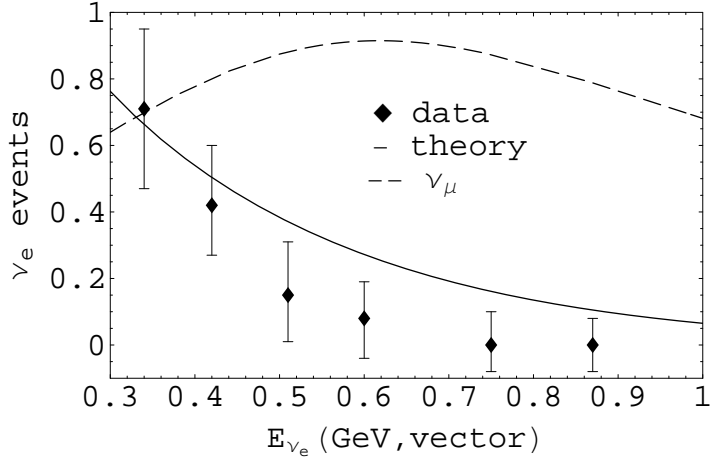}
\end{tabular}
\end{center}
\caption{ The energy spectrum for the decay of $\nu_\mu\to
\nu_e+\Un$ with vector unparticle where $d_\Un=2.05$,
$\Lambda_\Un=1$ TeV and $c_S=c_V=1$. } \label{fig6}
\end{figure}

Thirdly, we discuss the relative energy spectrum of final electron
neutrino. Fig. \ref{fig4} plots the initial $\nu_\mu$ energy
spectrum. The distribution is a quasi-Gaussian function, which peaks
around 700 MeV. If the muon neutrino decays to electron neutrino and
a conventional particle, the energy spectrum of electron neutrino
has the same distribution as that the muon neutrino beam. From the
data of excess events plotted Fig. \ref{fig5}, obviously, the
$\nu_e$ energy spectrum is not consistent with the data, no matter
for low energy or high energy. The experimental data show that the
excess of $\nu_e$ events is a decreasing function rather than a
Gaussian distribution in the energy range $0.3<E_{\nu_e}<1.0$ GeV.
This excludes the neutrino decays with conventional particles.
Unparticle is not a regular particle and has no a fixed mass. The
energy distribution of $\nu_e$ with unparticle being in the final
state is different from the initial $\nu_\mu$ spectrum and the final
energy distribution of $\nu_e$ depends on both effects. Combining
the $\nu_\mu$ energy spectrum (energy spreading of the muon neutrino
beam) and the differential ratios of neutrino decay given in Eqs.
(\ref{e1S},\ref{e1V}), the final electron neutrino energy
distribution is depicted in Figs. \ref{fig5} and \ref{fig6} for
scalar and vector unparticles, respectively. Within the
non-oscillation region, i.e., at low energy $0.3<E_{\nu_e}<0.45$
GeV, the theory prediction is well consistent with the experimental
data. For the energy range $0.5<E_{\nu_e}<1.05$ GeV, the theory
prediction is slightly larger than the data. It is noted that in the
above figures, only the relative size is estimated, the absolute
magnitude is very small.

\section{Discussions and Conclusions}

Motivated by the new measurement of the MiniBooNE Collaboration,
which observed an excess of electron-like events at low energy, it
motivates us to search for a possible mechanism beyond the SM to
explain this phenomenon, so the first idea which hits our mind is
that $\nu_2$ might decay into $\nu_1$ accompanied by some other very
light products. We should testify if this scenario can produce
results which are theoretically plausible and can explain the data.

There may be several possible modes, the first one is
$\nu_{2}\rightarrow \nu_1+\bar\nu_1+\nu_1$ which is a three-body
decay, the second one is $\nu_{2}\rightarrow \nu_1+a$ where $a$ is a
single boson particle, for example an axion etc., and the third one
is $\nu_2\rightarrow \nu_1+\Un$ where $\Un$ represents an
unparticle. All the three possibilities cannot be realized in the
framework of the SM, so new physics beyond the SM is necessary. The
first one was numerically estimated in this work and our results
indicate that the decay rate determined by the three-body decay mode
is too small and is ruled out immediately. The second mode is a
two-body decay, therefore the spectrum of the electron neutrinos is
discrete and it is not consistent with the measurement of the
MiniBooNE. Even though we consider the energy spreading of the
incident muon neutrino beam, the shape of the resultant electron
neutrino bump cannot be well understood in this scenario. Therefore
the third candidate is the most favorable. In this work, we work out
the formulations of neutrino decays within the framework of
unparticle physics. The formulations in the laboratory frame are
given for the first time.

The smallness of the decay width given by our numerical results
indicates that the unparticle scenario may not explain the excess of
electron neutrinos at low energy. The life time predicted in the
unparticle model is qualitatively consistent with the the
cosmological constraint \cite{Serpico} which is about $10^{17}$ sec.
By eq. (1), we know the suppression of ${\rm exp}(-t/10^{17})$ with
$t\sim 10^{-7}$ sec., would kill any possibility of observing a
decay event. The reasons are: (1) very tiny neutrino mass (2) very
weak interactions between the unparticle and neutrino. Since our
numerical results are consistent with the cosmology constraints and
the results by other authors \cite{Zhou}, we can be convinced that
the calculation is right, but the proposal does not work here.

Thus in this work, we definitely obtain a negative conclusion that
the peak of electron-neutrino at lower energy observed by the
MiniMoone collaboration cannot be explained by the neutrino decay.
On other aspect, the phenomenon is there and demands theoretical
explanations, so that we propose another scenario which might
overcome the aforementioned restrictions which forbid the appearance
of electron neutrinos to appear at low energy for the MiniBooNE
experiments. We will present the scenario in our next work.

\section*{Acknowledgments}

This work was supported in part by NNSFC under contract Nos.
10475042, 10745002 and 10705015 and the special foundation of the
Education Ministry of China.


\begin{thebibliography}{99}

\bibitem{MiniBooNE} The MiniBooNE Collaboration (A.A. Aguilar-Arevalo et al.),
 arXiv:0704.1500 [hep-ex].

\bibitem{LSND} C. Athanassopoulos {\it et al.}, Phys. Rev. Lett.
 {\bf 77}, 3082-3085 (1996); {\it ibid} {\bf 81}, 1774-1777 (1998).

\bibitem{MS} M. Maltoni and T. Schwetz, arXiv:0705.0107 [hep-ph].

\bibitem{Bodek} A. Bodek, arXiv:0709.4004 [hep-ph].

\bibitem{MRS} E. Ma, G. Rajasekaran and I. Stancu, Phys. Rev. D {\bf 61}, 071302
 (2000).

\bibitem{PPS} S. Palomares-Ruiz, S. Pascoli and T. Schwetz, JHEP {\bf 0509},
 048 (2005).

\bibitem{Georgi1} H. Georgi, arXiv:hep-ph/0703260.

\bibitem{BZ} T. Banks and A. Zaks, Nucl. Phys. B {\bf 196}, 189 (1982).

\bibitem{Georgi2} H. Georgi, arXiv:0704.2457 [hep-ph].

\bibitem{CKY} K. Cheung, W.-Y. Keung, T.-C. Yuan, arXiv:0704.2588 [hep-ph].

\bibitem{LZ} M. Luo, G. Zhu, arXiv:0704.3532 [hep-ph].

\bibitem{CG1} C.-H. Chen, C.-Q. Geng, arXiv:0705.0689 [hep-ph].

\bibitem{Liao} Y. Liao, arXiv:0705.0837 [hep-ph].

\bibitem{DY1} G.-J. Ding and M.-L. Yan, arXiv:0705.0794 [hep-ph].

\bibitem{ACG1} T.M. Aliev, A.S. Cornell and N. Gaur, arXiv:0705.1326 [hep-ph].

\bibitem{LW} X.-Q. Li and Z.-T. Wei, arXiv:0705.1821 [hep-ph].

\bibitem{LWW} C.-D. Lu and W. Wang and Y.-M. Wang, arXiv:0705.2909 [hep-ph].

\bibitem{Ste} M.A. Stephanov, arXiv:0705.3049 [hep-ph].

\bibitem{FRS} P.J. Fox and A. Rajaraman and Y. Shirman, arXiv:0705.3092 [hep-ph].

\bibitem{Gre} N. Greiner, arXiv:0705.3518 [hep-ph].

\bibitem{Dav} H. Davoudiasl, arXiv:0705.3636 [hep-ph].

\bibitem{CGM} D. Choudhury, D.K. Ghosh and Mamta, arXiv:0705.3637 [hep-ph].

\bibitem{CH} S.-L. Chen and X.-G. He, arXiv:0705.3946 [hep-ph].

\bibitem{ACG2} T.M. Aliev, A.S. Cornell and N. Gaur, arXiv:0705.4542 [hep-ph].

\bibitem{MR} P. Mathews and V. Ravindran, arXiv:0705.4599 [hep-ph].

\bibitem{Zhou} S. Zhou, arXiv:0706.0302 [hep-ph].

\bibitem{DY2} G.-J. Ding and M.-L. Yan, arXiv:0706.0325 [hep-ph].

\bibitem{CG2} C.-H. Chen, C.-Q. Geng, arXiv:0706.0850 [hep-ph].

\bibitem{LL2} Y. Liao and J.-Y. Liu, arXiv:0706.1284 [hep-ph].

\bibitem{BFRS} M. Bander, J.L. Feng, A. Rajaraman and Y. Shirman, arXiv:0706.2677
 [hep-ph].

\bibitem{Rizzo} T.G. Rizzo, arXiv:0706.3025 [hep-ph].

\bibitem{CKY2} K. Cheung, W.-Y. Keung and T.-C. Yuan, arXiv:0706.3155 [hep-ph].

\bibitem{GN} H. Goldberg and P. Nath, arXiv:0706.3898 [hep-ph].

\bibitem{CHT} S.-L. Chen, X.-G. He and Ho-Chin Tsai, arXiv:0707.0187 [hep-ph].

\bibitem{Zwicky} R. Zwicky, arXiv:0707.0677 [hep-ph].

\bibitem{KO} T. Kikuchi and N. Okada, arXiv:0707.0893 [hep-ph].

\bibitem{MG} R. Mohanta and A.K. Giri, arXiv:0707.1234 [hep-ph].

\bibitem{HW} C.-S. Huang and X.-H. Wu, arXiv:0707.1268 [hep-ph].

\bibitem{Lenz} A. Lenz, arXiv:0707.1535 [hep-ph].

\bibitem{CGF}  D. Choudhury and D.K. Ghosh,  arXiv:0707.2074 [hep-ph].

\bibitem{Li} X. Li, Z. Wei et al. in preparation.

\bibitem{PDG} W.-M. Yao, {\it et al.}, Particle Data Group, J. Phys. {\bf G33},
 1 (2006).

\bibitem{Serpico} P.D. Serpico, Phys. Rev. Lett. {\bf 98}, 171301
 (2007).

\bibitem{CHLTW} S.-L. Chen, X.-G. He, X.-Q. Li, H.-C. Tsai and Z.-T. Wei,
 arXiv:0710.3663 [hep-ph].


\end{thebibliography}
\end{document}